\documentclass[preprint,float]{JHEP3}
\usepackage{epsfig}
\usepackage{citesort}
\usepackage[T1]{fontenc}
\usepackage[latin1]{inputenc}
\usepackage{amssymb,amsmath}

\providecommand{\tabularnewline}{\\}

\title{
Transverse momentum resummation for Higgs boson produced via 
$b\bar{b}$ fusion at hadron colliders} 

\author{
Alexander Belyaev$^1$,  Pavel M. Nadolsky$^2$ and C.-P. Yuan$^1$
\\
$^1$ Department of Physics and Astronomy, Michigan State University, 
East Lansing, MI 48824, USA\\
$^2$ High Energy Physics Division, Argonne National Laboratory,
Argonne, IL 60439-4815, USA
}

\preprint{\vbox{\hbox{MSU-HEP-xxxxxx}}\\
	   \vbox{\hbox{ANL-HEP-PR-05-83}}} 
\date{\today}

\abstract{
  We study the impact of initial-state multiple parton
  radiation on transverse momentum $(q_T)$ distribution of Higgs boson
  produced via bottom quark fusion at hadron colliders. The shape of
  the resulting $q_T$ distribution 
  is affected by the  bottom-quark mass corrections
  and by the strong kinematical behavior of the bottom-quark parton 
  density. We account for both features in the full range of
  $q_T$. To do this, we  formulate the resummation calculation in a
  general-mass factorization (S-ACOT) scheme and introduce 
  a correction in the resummed-term to account for the effect from 
  large-$q_T$ kinematics of Higgs boson. 
  The results of this resummation
  are compared to fixed-order and PYTHIA predictions. 
}


\begin{document}

\def\eslt{\not\!\!{E_T}}
\def\mslash{\not\!\!{m}}
\def\to{\rightarrow}
\def\Phat{\hat{\Phi}}
\def\pT{$p_{T}$}
\def\bi{\begin{itemize}}
\def\ei{\end{itemize}}
\def\be{\begin{equation}}
\def\ee{\end{equation}}
\def\bea{\begin{eqnarray}}
\def\eea{\end{eqnarray}}
\def\te{\tilde e}
\def\tl{\tilde l}
\def\tu{\tilde u}
\def\ts{\tilde s}
\def\tb{\tilde b}
\def\tf{\tilde f}
\def\td{\tilde d}
\def\tQ{\tilde Q}
\def\tL{\tilde L}
\def\tH{\tilde H}
\def\tst{\tilde t}
\def\ttau{\tilde \tau}
\def\tmu{\tilde \mu}
\def\tg{\tilde g}
\def\tnu{\tilde\nu}
\def\tell{\tilde\ell}
\def\tq{\tilde q}
\def\tw{\widetilde W}
\def\tz{\widetilde Z}
\def\ttb{t \overline{t}}
\def\qqb{q \overline{q}}
\def\alt{\stackrel{<}{\sim}}
\def\agt{\stackrel{>}{\sim}}
\def\X{\times}
\def\emis{\not\hskip-5truedd E_{T} }
\def\LQ{{L\!\!Q}}
\def\tanb{\tan\beta}
\def\hbb{{\cal H}(b\bar{b})}
\def\mH{\mathcal{H}}

\def\toploop{{\ \rightarrow\hspace*{-0.4cm}^\rhd}\ }
\def\bbH {$b\bar{b}\rightarrow H$ }
\def\bbA {$b\bar{b}\rightarrow A$ }
\def\bbmH{$b\bar{b}\rightarrow \mH$ }
\def\ggH {$gg \toploop H$ }
\def\ggA {$gg\toploop A$ }
\def\ggmH{$gg\toploop \mH$ }

\section{Introduction}  

Understanding of electroweak symmetry breaking (EWSB)
is the central challenge for high energy physics. The search for Higgs
boson(s), assumed to be responsible for the generation of 
gauge-boson and fermion masses, is the primary task 
for the existing and future high energy colliders.

The Higgs sector may be represented by one complex 
scalar doublet, as it is economically realized in the Standard Model (SM),
or by two or more doublets, as it takes place 
in the Minimal Supersymmetric Standard Model (MSSM) 
and its extensions.
In MSSM, two Higgs-doublet superfields are necessary to 
generate masses for up- and down-type quarks
and  provide remarkable cancellation of triangle anomalies.

The two MSSM Higgs doublets have independent
vacuum expectation values (VEVs), $v_{u}$ and $v_{d}$. The sum of the
squares of these VEVs is fixed by the well-known $Z$ boson mass, but their
ratio, denoted as $\tan \beta =v_{u}/v_{d}$, is a free parameter of the
model. As a result of spontaneous symmetry breaking, five physical particles
appear in the Higgs sector: $h$ (light CP-even), $H$ (heavy CP-even), 
$A$ (CP-odd), and $%
H^{\pm }$ (charged).
An important feature of MSSM  is that,
for large values of $\tan \beta$, the Yukawa
couplings of the $b$-quarks to the neutral Higgs boson $\mH$ 
 (where $\mH = h$, $H$,  or $A$)
are enhanced by a factor $1/\cos \beta $ compared to 
the SM $b\bar b H_{SM}$ Yukawa coupling.

In the MSSM, Higgs boson masses are functions of CP-odd Higgs mass,
$m_{A}$,  and $\tan \beta $. 
The experimental lower limit on the Higgs boson mass
deduced from the LEP data~\cite{unknown:2001xx} favors scenarios 
with a intermediate-to-large values of $\tan \beta$ ($\gtrsim 5-10$). 
Theoretically, scenarios with large $\tan \beta $ are highly
motivated by SO(10) models of supersymmetric grand unification%
~\cite%
{Mohapatra:1999vv,Gell-Mann:1976pg,Fritzsch:1974nn,Raby:2004br,%
Altarelli:2004za,Ananthanarayan:1991xp,Anderson:1993fe,Carena:1994bv,%
Rattazzi:1995gk,Ananthanarayan:1994qt,Blazek:1999ue,Blazek:2001sb,%
Blazek:2002ta,Baer:1999mc,Baer:2000jj,Baer:2001yy,Auto:2003ys,Auto:2004km}.
Therefore, processes involving  Higgs boson production via enhanced 
$b\bar{b}\mH$ Yukawa coupling in the MSSM and other new physics models
may have large production rates  
and could play an important role 
in the study of the Higgs boson~\cite{Belyaev:2005nu}.

The  partonic processes contributing to the inclusive Higgs boson production
with enhanced $b\bar{b}\mH$ coupling are represented by 
(a) $b\bar{b}\rightarrow \mH$; (b) $gb\rightarrow \mH b $; 
and (c) $gg\rightarrow b\bar{b}\mH$ scattering.
In perturbative quantum chromodynamics (QCD), these are not independent
production mechanisms, since $b$ partons inside the hadron beam/target arise
from QCD evolution (splitting) of gluons, and gluons radiate off quarks \cite%
{Collins:1986mp,Olness:1988ub,Barnett:1987jw}. The three processes (a,b,c)
all give rise to the \emph{same hadronic final states}, with two $B$-mesons
appearing in different, but overlapping, regions of phase space---either
as beam/target remnants, or as high transverse momentum particles. The distinction
between the three processes depends very much on the factorization scheme
adopted for the QCD calculation, as has been 
recently reviewed in Ref.~\cite{Belyaev:2005nu}.

These {$\mH$}${b\bar{b}}$ processes have been extensively studied recently
in  SM and MSSM scenarios~\cite%
{Diaz-Cruz:1998qc,Balazs:1998nt,Dicus:1998hs,Campbell:2002zm,Maltoni:2003pn,%
Boos:2003yi,Hou:2003fm,Dittmaier:2003ej,Harlander:2003ai,Dawson:2003kb,%
Kramer:2004ie,Field:2004nc,Dawson:2004sh,Dawson:2005vi}. 
Calculations in the 5-flavor scheme have been carried out to the 2-loop
level \cite{Harlander:2003ai}, which considerably reduced the theoretical
uncertainty due to the perturbative expansion, as estimated by the residual
scale dependence. \ Comparison of results obtained in the 4- and 5-flavor
schemes has also been carried out \cite{Dawson:2004sh,Dawson:2005vi}. 
It shows consistency
between the two schemes in the energy region of the 
Fermilab Tevatron and the 
 CERN Large Hadron Collider~(LHC).

In spite of the good theoretical control of
the inclusive Higgs boson production cross section
in $b\bar{b}$ fusion, neither the 4-flavor nor 5-flavor 
scheme is adequate for
predicting the transverse momentum ($q_T$) distribution
of the Higgs boson, when $q_T$ values are of the order of the 
$b$ quark mass ($m_b$), 
at any fixed order of the perturbative calculation.
To properly describe the low-$q_T$ region and determine the
$q_T$ range where fixed-order calculations are applicable,
one should  resum large soft and collinear logarithms, as we 
discuss in detail in the next section.
Furthermore, a more
comprehensive factorization scheme (general-mass variable flavor
number scheme) must be used to describe non-negligible dependence on
$m_b$ when $q_T$ is comparable to $m_b$.
Many studies of soft gluon resummation 
for the kinematical distributions of Higgs boson produced via
gluon-gluon fusion are available in the literature~\cite{%
Yuan:1991we,Berger:2002ut,Catani:2003zt,Kulesza:2003wn,Gawron:2003np,%
Field:2004tt,Grazzini:2004km,Bozzi:2005wk}.
As for the soft gluon resummation in
production of Higgs boson via $b\bar b$ fusion, 
it was first discussed for massless $b$ quarks in Ref.~\cite{Balazs:1998sb}.
Later, Ref.~\cite{Field:2004nc} has 
studied the effect of $O(\alpha_s^2)$
subleading logarithmic contribution in the Sudakov form factor, 
but only in the soft-gluon limit.

As shown in Refs.~\cite{Belyaev:2002zz,Belyaev:2005ct}, the
correct model for the transverse momentum distribution 
of the Higgs boson is crucial 
for unambiguous reconstruction of Higgs boson mass
in the $\mH\to\tau\tau$ decay channel. 
It is also important 
for discriminating the signal events from the backgrounds 
by carefully examining the $q_{T}$ distribution of the Higgs boson  
in ${\mH}b\bar{b}$ associated production,
followed by ${\mH}\to b\bar{b}$ decay~\cite{Carena:2000yx}.
In this work, we 
study the effect of the initial-state multiple parton radiation
on the transverse momentum distribution of Higgs boson produced at hadron
colliders via $b\bar b$ fusion.
Our calculation includes  
contributions from resummation up to next-to-next-to-leading logarithmic 
 accuracy and fixed-order perturbation theory up to next-to-leading
 order.\footnote{ 
This order of accuracy is the same as that in
the calculation~\cite{Balazs:1997xd} of
the kinematical distributions for 
weak gauge bosons produced via light quark annihilation in hadron collisions.
}
In view of the non-negligible mass of the bottom quark and strong
kinematical dependence of the $b$-quark parton density,
special corrections (referred in this work 
as the ``heavy-quark mass correction'' and ``kinematical correction'')
must be included in the resummation formalism
to predict the $b\bar b \rightarrow \mH$ cross section
in the full range of $q_{T}$.

The rest of this paper is organized as follows.
In Section 2, we discuss the theoretical framework for  
transverse momentum resummation used in this study and introduce 
the ``heavy-quark mass correction'' and
``kinematical correction'''. 
In Section 3, we present the numerical results for $b\bar b \rightarrow
\mH$ at the Tevatron and the LHC and compare the $q_T$
distributions predicted by resummation calculations to 
PYTHIA predictions. Section 4 contains our conclusions.

\section{Resummation for heavy flavors: theory\label{sec:InclXS}}

Kinematical distributions of Higgs boson produced in bottom quark-antiquark
fusion at hadron colliders are often computed within the zero-mass
variable flavor number (ZM-VFN) factorization scheme, which neglects
masses of bottom and lighter quarks in hard-scattering amplitudes.
The 5-flavor scheme is a realization of the ZM-VFN scheme at
energies above the bottom-quark mass threshold.
The application of this scheme is justified because the mass $M_{\mH}$
of the Higgs boson of interest lies in the range of a few hundred
GeV, \emph{i.e.}, it is much larger than the mass $m_{b}$ of the
bottom quark. If all momentum scales in the cross section are of order
$Q\sim M_{\mH}\gg m_{b}$, for instance, when the inclusive rate for
Higgs boson production is computed, the mass dependence of the hard-scattering
amplitude is suppressed by powers of a small factor $m_{b}^{2}/Q^{2}$
and can be safely discarded. Meanwhile, large logarithms $\ln^{p}(Q^{2}/m_{b}^{2})$,
arising from $b\bar{b}$ pairs being 
produced in collinear splittings
of gluons, are resummed in 
the parton distribution function (PDF) of $b$-quark
in the proton by utilizing Dokshitzer-Gribov-Lipatov-Altarelli-Parisi
(DGLAP) equations \cite{Gribov:1972ri,Gribov:1972rt,Dokshitzer:1977sg,Altarelli:1977zs}.
The perturbative parton distribution of the bottom quark is set identically
equal to zero at factorization scales $\mu_{F}$ below the $b$-quark
mass threshold ($\mu_{F}<m_{b}$) and turned on at $\mu_{F}\geq m_{b}$.
In technical terms, such a definition corresponds to regularization
of the ultraviolet divergence in the perturbative $b$-quark PDF by
zero-momentum subtraction at $\mu_{F}<m_{b}$ and $\overline{MS}$
subtraction at $\mu_{F}\geq m_{b}$ \cite{Collins:1978wz}.

The 4-flavor scheme can be viewed as a realization of the fixed-flavor
number, or FFN, scheme
\cite{Gluck:1982cp,Gluck:1988uk,Nason:1989zy,Laenen:1992cc,Laenen:1993zk,Laenen:1993xs},
as far as the bottom quarks are concerned.
This scheme 
does not introduce nontrivial parton densities for heavy quarks, but
rather keeps the heavy-quark contributions, including the logarithmic
terms $\ln^{p}(Q^{2}/m_{b}^{2})$, in the hard-scattering amplitudes.
The heavy-quark PDF is set identically to zero in the FFN scheme,
as a consequence of regularization by zero-momentum subtraction in
the whole range of $\mu_{F}$. As in the ZM-VFN scheme, the mass-dependent
terms are negligible in the FFN scheme at $Q^{2}\gg m_{b}^{2}$, except
for the logarithms $\ln^{p}(Q^{2}/m_{b}^{2})$, which, however, are
not resummed into the heavy-quark parton densities. 
Hence, at large $Q$, the factorization scale dependence in the
FFN perturbative calculation is stronger than that in the ZM-VFN
calculation, though the FFN calculation is more reliable when 
$Q$ is close to $m_{b}$. 
Both ZM-VFN and FFN schemes were applied recently
to Higgs boson production via $b\bar{b}$ fusion with zero, one or
two tagged $b$ quarks in the final state \cite{Harlander:2003ai,Campbell:2002zm,Maltoni:2003pn,Dawson:2003kb,Dawson:2004sh,Dawson:2005vi,Dittmaier:2003ej}.
A recent computation of $O(\alpha_{s}^{2})$ corrections \cite{Harlander:2003ai}
to the $b\bar{b}\rightarrow \mH$ cross section was realized in the
ZM-VFN scheme, since implementation of the quark mass dependence is
tedious beyond $O(\alpha_{s})$.

However, neither the ZM-VFN scheme nor the FFN scheme will work well
in a calculation of the transverse momentum distribution 
of the Higgs boson, when $q_{T}$ is small, of the order
of the bottom quark
mass. To properly describe the small-$q_{T}$ region,
large soft and collinear logarithms $\ln^{m}(q_{T}^{2}/Q^{2})$ must
be resummed together with the mass logarithms $\ln^{p}(Q^{2}/m_{b}^{2})$,
while keeping the essential dependence on the $b$-quark mass at $q_{T}\approx m_{b}$.
Soft gluon resummation must be realized in a more comprehensive factorization
scheme (a general-mass variable flavor number scheme \cite{Collins:1998rz})
to achieve this goal \cite{Nadolsky:2002jr}. Formulation of the Collins-Soper-Sterman
(CSS) resummation method \cite{Collins:1985kg} in such a scheme (simplified
Aivazis-Collins-Olness-Tung, or S-ACOT, scheme \cite{Collins:1998rz,Kramer:2000hn})
has been proposed recently \cite{Nadolsky:2002jr} to compute $q_{T}$
distributions in scattering processes initiated by heavy quarks. The
advantage of the S-ACOT scheme is that it retains massless expressions
for hard-scattering amplitudes with incoming heavy quarks (flavor-excitation
amplitudes), and hence, drastically simplifies the calculation, while preserving
the relevant mass terms. In this work, we adopt the $q_{T}$ resummation
formalism with the heavy-quark (HQ) mass effects proposed in Ref.~\cite{Nadolsky:2002jr},
hereafter referred to as {}``the CSS-HQ formalism''.

The CSS formalism and its application to $b\bar{b}\rightarrow \mH$
in the ZM-VFN scheme has been discussed extensively in Refs.~\cite{Balazs:1997xd,Balazs:1998sb,Landry:2002ix},
and we refer the reader to those publications for details. Here we
summarize the features of the CSS formalism essential for the understanding
of the numerical results in the next section. We symbolically write
the total (TOT) resummed differential cross section $d\sigma/dQ^{2}dydq_{T}^{2}$
for $h_{1}(P_{1})h_{2}(P_{2})\stackrel{b\bar{b}}{\longrightarrow}\mH (q)X,$
where $h_{1}$ and $h_{2}$ are the initial-state hadrons, as\begin{equation}
\mbox{TOT}=\mbox{W}+\mbox{PERT}-\mbox{ASY}.\label{TOT}\end{equation}
 ${\rm W}$ denotes the Fourier-Bessel integral of a resummed form factor
$\widetilde{W}(b,Q,x_{1},x_{2})$ introduced in space of the impact
parameter $b$ (Fourier-conjugate to $q_{T}$):\[
{\rm W}\equiv\frac{1}{(2\pi)^{2}}\int_{0}^{\infty}d^{2}b\, e^{i\vec{q}_{T}\cdot\vec{b}}\,\widetilde{W}(b,Q,x_{1},x_{2}).\]
 PERT is the perturbative QCD cross section, evaluated at a finite
order of the QCD coupling strength $\alpha_{s}$. The asymptotic piece
ASY is defined as a perturbative QCD expansion of the resummed ${\rm W}$-term
to the same order of $\alpha_{s}$ as in PERT. The difference
of PERT and ASY, which we denote by  ${\rm Y}\equiv {\rm PERT} - {\rm ASY}$, is 
the regular part of the cross section. $x_{1}$ and $x_{2}$ are the light-cone momentum
fractions of the partons entering the $b\bar{b}\mH$ vertex in ${\rm W}$
and ASY. They satisfy $x_{1,2}=Qe^{\pm y}/\sqrt{S}$ at $q_{T}\rightarrow0$,
with $Q$ and $y=\frac{1}{2}\ln((q^0+q^3)/(q^0-q^3))$ being the invariant mass and rapidity of the Higgs
boson, and the square of the $h_1h_2$ center-of-mass energy 
$S\equiv(P_{1}+P_{2})^{2}$. We will discuss the choice
of $x_{1,2}$ at nonzero $q_{T}$ in a moment. 

The resummed form factor $\widetilde{W}(b,Q,x_{1},x_{2})$ is composed
of perturbative QCD contributions (dominant at $b\rightarrow0$) and
non-perturbative contributions (dominant at $b\rightarrow\infty$).
We adopt the $b_{*}$ prescription \cite{Collins:1982va,Collins:1985kg}
to evaluate the form factor $\widetilde{W}(b,Q,x_{1},x_{2})$ 
numerically, which is given
by\begin{equation}
\widetilde{W}(b,Q,x_{1},x_{2})=\widetilde{W}_{pert}(b_{*},Q,x_{1},x_{2})e^{-{\cal F}_{NP}(b,Q)}.\label{Wfull}\end{equation}
In Eq.~(\ref{Wfull}), $\widetilde{W}_{pert}(b_{*},Q,x_{1},x_{2})$
is the perturbative form factor, evaluated as a function of the variable
$b_{*}\equiv b/\sqrt{1+b^{2}/b_{max}^{2}}$ with $b_{max}$ being
a free parameter:\begin{eqnarray}
\widetilde{W}_{pert}(b,Q,x_{1},x_{2}) & = & \frac{\pi}{S}\,\sum_{a_{1},a_{2}}\sigma^{(0)}\, e^{-\mathcal{S}_{P}(b,Q)}\,\,\label{WCSS}\\
 &  & \left[({\cal C}_{b/a_{1}}\otimes f_{a_{1}/h_1})(x_{1},b)\,\,({\cal
 C}_{\bar{b}/a_{2}}\otimes f_{a_{2}/h_2})(x_{2},b)+(b\leftrightarrow\bar{b})\right].\end{eqnarray}
 Here \begin{eqnarray}
{\mathcal{S}}_{P}(b,Q)\equiv\int_{C_{1}^{2}/b^{2}}^{C_{2}^2 Q^{2}}\frac{d\bar{\mu}^{2}}{\bar{\mu}^{2}}\biggl[{\mathcal{A}}(\alpha_{s}(\bar{\mu}))\,\mathrm{ln}\biggl(\frac{Q^{2}}{\bar{\mu}^{2}}\biggr)+{\mathcal{B}}(\alpha_{s}(\bar{\mu}))\biggr]\label{Sudakov}\end{eqnarray}
is the perturbative Sudakov factor, with $b_{0}\equiv2e^{-\gamma_{E}}\approx1.123$,
and

\begin{equation}
\sum_{a}({\cal C}_{b/a}\otimes f_{a/h})(x,b)\equiv\int_{x_{1}}^{1}\frac{d\xi}{\xi}{\cal C}_{b/a}(\xi,b,\mu_{F},m_{b})f_{a/h}(\frac{x}{\xi},\mu_{F})\label{CxF}\end{equation}
 is a convolution of the Wilson coefficient function ${\cal C}_{b/a}(x,b,\mu_{F},m_{b})$
for outgoing $b$ quarks and parton density $f_{a/h}(x,\mu_{F}),$
summed over the intermediate parton states $a$. The overall normalization
is given by \begin{equation}
\sigma^{(0)}=\frac{\pi}{6}\left(
\frac{m_b(Q)~\tan\beta}{v}\right)^{2}
\delta(Q^2-M_{\mH}^2) \,,\label{sigma0}
\end{equation}
 where
$\beta$ and $v$ are the Higgs-doublet mixing angle and vacuum expectation
value, respectively. The bottom quark 
running mass $m_b(Q)$ is evaluated at the scale $Q$
\cite{Balazs:1998sb},
and $M_{\mH}$ is the mass of the Higgs boson $\mH$. 

The non-perturbative contributions are described by the exponential
$e^{-{\cal F}_{NP}(b,Q)}$. For simplicity, the function ${\cal F}_{NP}(b,Q)$
is assumed to be the same in the heavy- and light-quark channels and
taken from the global $q_{T}$ fit \cite{Landry:2002ix} for $b_{max}=0.5\mbox{ GeV}^{-1}$.
This approximation is sufficient for the purposes of the comparison
of the resummation calculation with PYTHIA and may be refined by using
alternative prescriptions, such as the revised $b_{*}$ model \cite{Konychev:2005iy},
in the future analyses.

The perturbative expressions for the functions PERT, ASY, and 
the coefficients ${\mathcal{A}}^{(1)}$, ${\mathcal{B}}^{(1)}$  
and ${\mathcal{A}}^{(2)}$ of the function ${\rm W}$
can be found in Refs.~\cite{Balazs:1997xd,Balazs:1998sb}.
The ${\cal O}(\alpha_{s}/\pi)$ coefficients ${\cal C}_{b/a}^{(1)}(x,b,\mu_{F})$
in the Wilson coefficient functions ${\cal C}_{b/a}(x,b,\mu_{F})$
are different in the ZM-VFN and S-ACOT factorization schemes.
In the ZM-VFN scheme, 
\begin{eqnarray}
{\cal C}_{b/b}^{(1)}(x,b,\mu_{F}) & = & \frac{C_{F}}{2}(1-x)-\ln\left(\frac{\mu_{F}b}{b_{0}}\right)P_{q/q}^{(1)}(x)\nonumber \\
 & + & C_{F}\delta(1-x)\left[-\ln^{2}\left(\frac{C_{1}}{b_{0}C_{2}}e^{-3/4}\right)+\frac{\cal V}{4}+\frac{9}{16}\right],\label{eq:C01}\\
{\cal C}_{b/g}^{(1)}(x,b,\mu_{F}) & = & T_{R}x(1-x)-\ln\left(\frac{\mu_{F}b}{b_{0}}\right)P_{q/g}^{(1)}(x),\label{eq:C02}\end{eqnarray}
where $T_{R}=1/2$ and $C_{F}=(N_c^2-1)/(2N_c)=4/3$ 
are the QCD Casimir invariants (with $N_c=3$),
$P_{q/q}^{(1)}(x)=C_{F}[(1+x^{2})/(1-x)]{}_{+}$ and $P_{q/g}^{(1)}(x)=T_{R}(x^{2}+(1-x)^{2})$
are the ${\cal O}(\alpha_{s})$ splitting functions, 
and ${{\cal V}}=$$\pi^{2}-2$.%
\footnote{The ${\cal V}$ term shown here is different from that in Eq.~(14)
in Ref.~\cite{Balazs:1998sb}, as a result of a typo in
Eq.~(14), cf. Eq.~(A2), of that paper.}

The coefficient ${\mathcal{B}}^{(2)}$ for $N_f$ active quark flavors 
can be obtained following the
method presented in Refs.~\cite{deFlorian:2000pr,deFlorian:2001zd}. 
It is found to be 
\begin{equation}
{{\cal B}}^{(2)}={{\cal B}}_{universal}^{(2)}+\beta_{0}\frac{{{\cal V}}}{4}
\end{equation}
with $\beta_{0}=(11N_{c}-2N_{f})/6$,
\begin{eqnarray}
{{\cal B}}_{universal}^{(2)} & = & -\frac{\delta P_{q/q}^{(2)}}{2}+
\frac{\beta_{0}C_{F}\pi^{2}}{12}+\beta_{0}C_{F}
\left(\left(\ln\frac{b_{0}}{C_{1}}\right)^{2}-\frac{3}{2}
\ln C_{2}-\ln^{2}C_{2}\right)\nonumber\\
 & - & C_{F}\left(\left(\frac{67}{36}-\frac{\pi^{2}}{12}\right)N_c-
\frac{5}{9}T_{R}N_{f}\right)\ln\left(\frac{b_{0}^{2}C_{2}^{2}}{C_{1}^{2}}
\right),
\end{eqnarray}
and
\begin{equation}
\delta P_{q/q}^{(2)}=C_{F}^{2}\left(\frac{3}{8}-
\frac{\pi^{2}}{2}+6\zeta_{3}\right)+C_{F}N_c\left(\frac{17}{24}+
\frac{11\pi^{2}}{18}-3\zeta_{3}\right)-C_{F}T_{R}N_{f}\left(
\frac{1}{6}+\frac{2\pi^{2}}{9}\right) \,,
\end{equation}
where  $\zeta_3 = 1.202057...$ is the Riemann zeta-function $\zeta_n$ for $n=3$. 
Throughout the paper, we choose the factorization constants $C_{1}=b_{0}$
and $C_{2}=1,$ and the factorization scale $\mu_{F}=b_{0}/b$ in
the Wilson coefficient functions ${\cal {\cal C}}_{b/a}(x,b,\mu_{F})$
and parton distributions $f_{a/h}(x,\mu_{F}).$
We note that the above results for the
${\cal C}^{(1)}$ and ${\mathcal{B}}^{(2)}$ coefficients 
are different from the ones presented in 
Ref.~\cite{Field:2004nc}, where only soft gluon contributions
 were considered.
 
In the S-ACOT scheme, we include $m_{b}$ dependence in the Wilson
coefficient ${\cal C}_{b/g}^{(1)}(x,b,\mu_{F})$ for gluon splitting into
a $b \bar b$ pair (\emph{i.e.,}  $b\leftarrow g$
splittings):\begin{eqnarray}
{\mathcal{C}}_{b/g}^{(1)}(x,b,m_{b},\mu_{F}) & = & T_{R}x(1-x)\, b\, m_{b}\, K_{1}(b\, m_{b})\nonumber \\
 &  & +P_{q/g}^{(1)}(x)\left[K_{0}(b\, m_{b})-\theta(\mu_{F}-m_{b})\ln\Bigl(\frac{\mu_{F}}{m_{b}}\Bigr)\right],\label{CHQg}\end{eqnarray}
 where $K_{0}(z)$ and $K_{1}(z)$ are the modified Bessel functions~\cite{AbramowitzStegun}.
This expression reduces to the massless result in Eq.~(\ref{eq:C02})
when $b\ll m_{b}$:\begin{equation}
\lim_{b\, m_{b}\rightarrow0}{\mathcal{C}}_{b/g}^{(1)}(x,b,m_{b},\mu_{F})=
T_{R}x(1-x)-\ln\left(\frac{\mu_{F}b}{b_{0}}\right)P_{q/g}^{(1)}(x)
\,,
\label{CHQgmeq0}\end{equation}
since $K_{0}(z) \rightarrow 0 $ and $K_{1}(z) \rightarrow 1/z$ as 
$z\rightarrow 0$.
We keep massless expressions for the remaining perturbative coefficients
in W, PERT, and ASY, in accordance with the rules of the S-ACOT
scheme~\cite{Nadolsky:2002jr}. In particular, the mass dependence
is neglected in PERT and ASY, as their difference contributes substantially
to Eq.~(\ref{TOT}) only at $q_{T}\sim Q\gg m_{b}$. 

While PERT is evaluated by using the exact $2\rightarrow 2$ kinematics,
the phase space element in the ${\rm W}$ and ASY terms is unique only in
the limit $q_{T}\rightarrow0$, but may be defined in several ways
at nonzero $q_{T}$. At $q_{T}$ of order $Q,$ momentum fractions
$x$ of the initial-state partons must be large enough to produce
the Higgs boson with a large transverse mass $M_{t}=\sqrt{Q^{2}+q_{T}^{2}}$;
too small momentum fractions $x\sim Q/\sqrt{S}$ accessible at $q_{T}\rightarrow0$
are not allowed. To introduce information about the reduction of phase
space available for collinear QCD radiation at large $q_{T}$, we
define the light-cone momentum fractions $x_{1}$ and $x_{2}$ in
${\rm W}$ and ASY as 
\begin{equation}
x_{1,2}\equiv2(P_{2,1}\cdot q)/S=M_{t}e^{\pm y}/\sqrt{S} \,.
\label{x12}
\end{equation}
As $q_{T}\rightarrow0$, $x_{1}$ and $x_{2}$ reduce to their Born-level
expressions, $x_{1,2}^{\mbox{Born}}=Qe^{\pm y}/\sqrt{S}$, and the
canonical CSS form is reproduced. At larger $q_{T}$, contributions
from unphysical small momentum fractions are excluded from ${\rm W}$ and
ASY by the growing $M_{t}$ in $x_{1,2}$. We will argue that this 
prescription of redefining the values of $x_{1,2}$, to be referred 
as the {}``kinematical correction'' in this work, is
crucial for predicting the resummation cross sections in $b\bar{b}$
channel at large $q_{T}$, due to the strong dependence of the
$b$-quark PDF on $x$ in the relevant kinematic regions of 
the Tevatron and the LHC. 

The next section presents numerical comparisons of the parts of the
resummation cross sections (TOT, ${\rm W}$, PERT, and ASY), computed under
various assumptions about the order of the perturbative coefficients
in $\alpha_{s}$, factorization scheme, and kinematical correction.
Our naming conventions for various terms are summarized in Table~\ref{table:Naming-conventions}.
A numerical argument in the names of the PERT and ASY functions indicates
the order of the perturbative calculation. For example, PERT(1) stands
for the ${\cal O}(\alpha_{s})$ perturbative piece. Three arguments
following the names of the functions ${\rm W}$ and TOT indicate orders
of the functions ${\cal A}$ and ${\cal B}$ in the Sudakov factor
(\ref{Sudakov}), and Wilson coefficient functions ${\cal C}_{b/a}(x,b,\mu_{F},m_{b})$
in Eq.~(\ref{CxF}). For example, the ${\cal A}$, ${\cal B}$, and
${\cal C}$ functions in the W(2,2,1) distribution are evaluated up
to orders $\alpha_{s}^{2}$, $\alpha_{s}^{2}$, and $\alpha_{s}$,
respectively. The S-ACOT factorization scheme and active kinematical
correction are indicated by the superscript {}``HQ'' (for {}``heavy-quark'')
and subscript {}``KC'' (for {}``kinematical correction''), as in W${}_{\mbox{KC}}^{\mbox{HQ}}$(1,1,0)
for the ${\rm W}(1,1,0)$-term with the kinematical correction evaluated in the
S-ACOT scheme. 

\TABLE{
\begin{tabular}{
|c|c|c|c|c|c|c|}
\hline 
&
\multicolumn{4}{c|}{Order of QCD coupling strength $\alpha_{s}$}&
Kinematical&
S-ACOT\tabularnewline
\cline{2-5} 
&
${{\cal A}}(\alpha_{s}(\bar{\mu}))$&
${{\cal B}}(\alpha_{s}(\bar{\mu}))$&
${{\cal C}}_{b/a}$&
${\rm Y}$&
correction&
scheme\tabularnewline
\hline
\hline 
{\small TOT(1,1,0)}&
1&
1&
0&
1&
-&
-\tabularnewline
\hline 
{\small TOT(2,2,1)}&
2&
2&
1&
1&
-&
-\tabularnewline
\hline 
{\small TOT${}_{\mbox{KC}}$(2,2,1)}&
2&
2&
1&
1&
yes&
-\tabularnewline
\hline
\hline 
{\small W(1,1,0)}&
1&
1&
0&
-&
-&
-\tabularnewline
\hline
{\small W(2,2,1)}&
2&
2&
1&
-&
-&
-\tabularnewline
\hline
{\small W${}_{\mbox{KC}}$(2,2,1)}&
2&
2&
1&
-&
yes&
-\tabularnewline
\hline
{\small W${}_{\mbox{KC}}^{\mbox{HQ}}$(2,2,1)}&
2&
2&
1&
-&
yes&
yes\tabularnewline
\hline
\hline 
{\small PERT(1)}&
-&
-&
- &
1&
-&
-\tabularnewline
\hline
{\small ASY(1)}&
-&
-&
-&
1&
-&
-\tabularnewline
\hline
{\small ASY${}_{\mbox{KC}}$(1)}&
-&
-&
-&
1&
yes&
-\tabularnewline
\hline
\end{tabular}
\caption{\label{table:Naming-conventions} Naming conventions for $q_{T}$
distributions in Section~\protect\ref{sec:Numerical-results}.}
}

\section{Numerical results\label{sec:Numerical-results}}

In this section we present the numerical results of our study for 
a Higgs boson  produced via bottom quark fusion at the
Tevatron Run-2 (a 1.96 TeV proton-antiproton collider)
and the LHC (a 14 TeV proton-proton collider).
To make our study less dependent on SUSY parameters, we focus on production of
the CP-odd Higgs particle $A$, since the Yukawa
couplings of $A$ to the heavy quarks are independent of the Higgs mixing angle.

First, we study in details the Tevatron case,
and then present our final results also for the LHC.
As an example, we have chosen $m_A$ to be 
100~GeV and 300~GeV, respectively, for the Tevatron and LHC studies.
Our results could be trivially generalized
to the production of any neutral Higgs boson $\mH$
by properly scaling the $b\bar{b}A$ Yukawa coupling to the actual one.
In our studies the value of $\tan \beta$ was chosen  to be 50.
The
tree level Yukawa coupling of bottom quarks and the CP-odd Higgs boson ($A$)
in the minimal supersymmetric Standard Model (MSSM)
is equal to 
${m_b \tan \beta/(\sqrt{2}v)}$, where the bottom quark mass is 4.7 GeV,
and the vacuum expectation value $v$ is about 246 GeV.
To improve the numerical predictions, we have resummed the large
logarithms originated from the QCD radiative corrections  
to the $b\bar{b}A$  Yukawa coupling by introducing the
running bottom quark mass at the scale of Higgs boson mass,
as done in Ref.~\cite{Balazs:1998sb} for calculating the next-to-leading order
QCD corrections to the production of $b \bar b \rightarrow A$ in hadron
collisions. The running bottom quark mass is about 2.98 GeV for 
$m_A=100$ GeV.

The full event generator, such as PYTHIA, can give a fair description of
the event topology after turning on the QCD showering from the initial
and, possibly, final states, which are generated by the probability
functions calculated from the relevant Sudakov form factors 
\cite{Sjostrand:2001yu}.
Therefore, it is a common practice to compare  PYTHIA predictions
 to the event
shape of experimental data. For example, the shape of the 
$q_T$ distribution of the 
vector boson produced at hadron colliders can be fairly described by
the PYTHIA program, though the normalization of the event rate is usually
off the scale, because it does not include 
all the finite part of higher order QCD corrections \cite{Balazs:1997xd}.
As compared to the CSS resummation formalism, the PYTHIA calculation
does not include contributions generated from the $C$-functions and the
Y-term.
Therefore, a fair comparison to the PYTHIA prediction is to include only the
Sudakov contributions in a resummation calculation. 
In Fig.~\ref{fig:one}, we show the prediction of W(1,1,0) and compare
it to the PYTHIA prediction for the hard scattering
process $b {\bar b} \rightarrow H $~\cite{Sjostrand:2001yu}.\footnote{
The invariant mass cutoff  of parton showers 
(the parameter PARJ(82) in the computer
code), below which partons are assumed not to radiate,
was chosen to be 1.0 GeV,
which is the default value of PYTHIA version 6.2.
}
It is evident that W(1,1,0) predicts a very different shape of 
$q_T$ distribution from
PYTHIA in $b \bar b \rightarrow A$ production, though 
the integrated rates (\emph{i.e.,} the areas under the two curves) 
are about the same. 
On the contrary, in the case of
vector boson production that is dominated by light quark scatterings in
the initial state, 
the above two calculations predict similar, though not
identical, $q_T$ distributions~\cite{Balazs:1997xd}. 
Fig.~\ref{fig:one} illustrates that the 
peak position of $q_T$ distribution
predicted by PYTHIA is lower than that by the resummation calculation 
 W(1,1,0). Also, PYTHIA predicts a narrower shape in 
the $q_T$ distribution.
It is important to understand the cause of this difference in order to
reliably predict the $q_T$ distribution of the Higgs boson produced 
in $b \bar b \rightarrow A$ process.

%
\FIGURE[htb]{
\noindent
\includegraphics[width=0.6\textwidth]{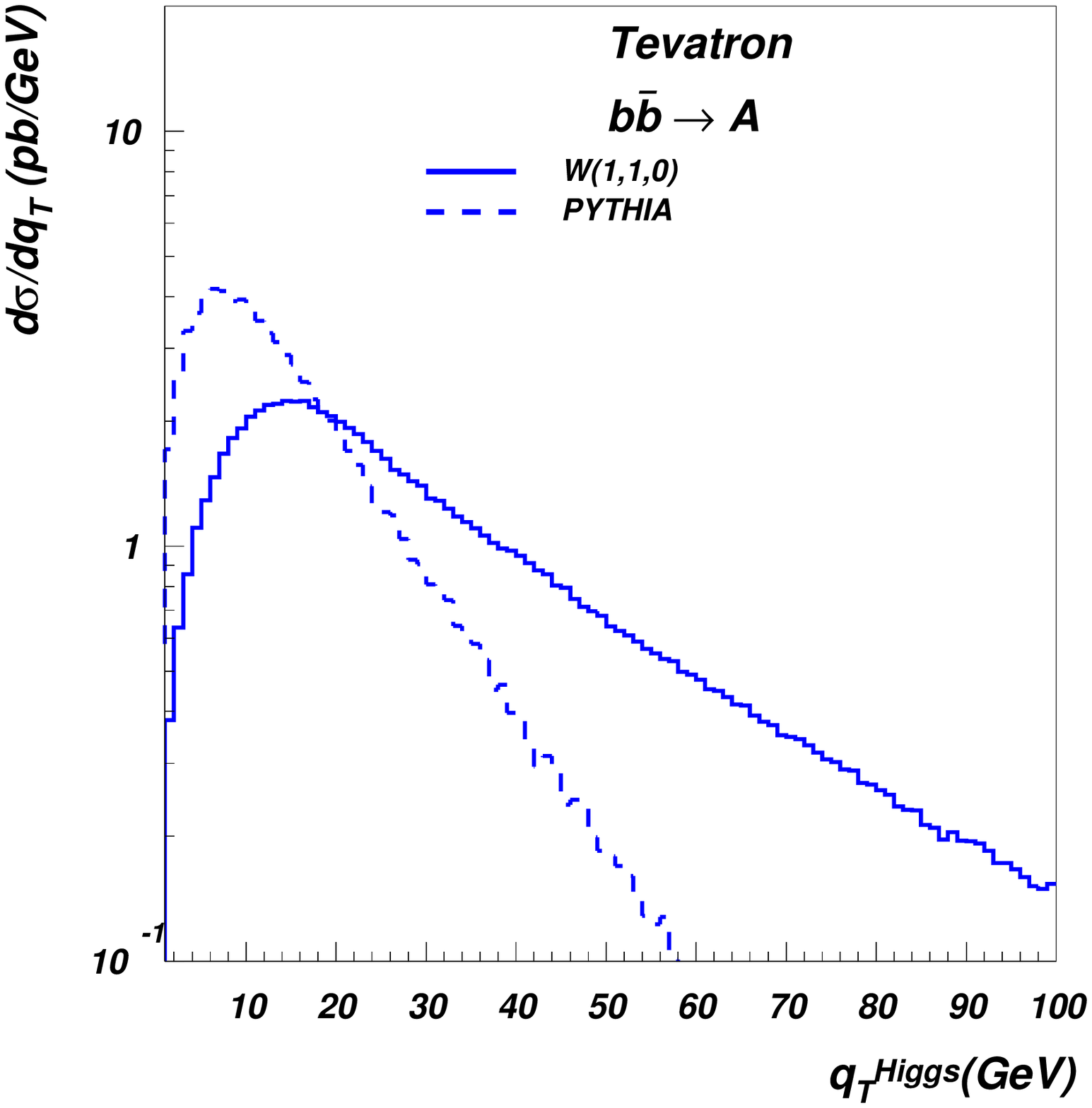}
\caption{
$q_T$ distributions predicted by ${\rm W}(1,1,0)$ and PYTHIA,
shown as solid and dashed curves, respectively, for a 100 GeV Higgs
boson produced via $b \bar b$ fusion at the Tevatron Run-2.}
\label{fig:one}
}

A close examination  reveals that when PYTHIA generates QCD showering,
the kinematical distributions of the final-state particles (including
the quarks and gluons generated from QCD showering) 
are slightly modified to satisfy
energy-momentum conservation at each stage of showering. Namely, the
kinematics of Higgs boson is modified according to the amount of showering. 
By this, it effectively includes some part of higher-order contribution
(similar to part of PERT(1) contribution). On the other hand, in the W(1,1,0)
calculation the emitted soft gluons are assumed not to carry any
momentum (strictly in the soft limit), which
 implies that the Y-term contribution
could be important for determining the shape of $q_T$ distribution. 
From Fig.~\ref{fig:one}, we expect the Y-term contribution
to be negative at large $q_T\sim m_A$ in order for 
the result of resummation calculation to resemble more the PYTHIA
prediction. Moreover, the Sudakov form factors in PYTHIA
contain some additional $O(\alpha_s^2)$ contributions when using the 
next-to-leading order PDF's to generate the event distributions.
Thus, we should improve the
resummation calculation by including next-to-leading order perturbative 
corrections. The resulting distribution, denoted as TOT(2,2,1),
is the sum of W(2,2,1) and
PERT(1) with the subtraction of ASY(1) to avoid the overlapped
contribution, cf. Eq.~(\ref{TOT}).

%
\FIGURE[htb]{
\noindent
\includegraphics[width=0.6\textwidth]{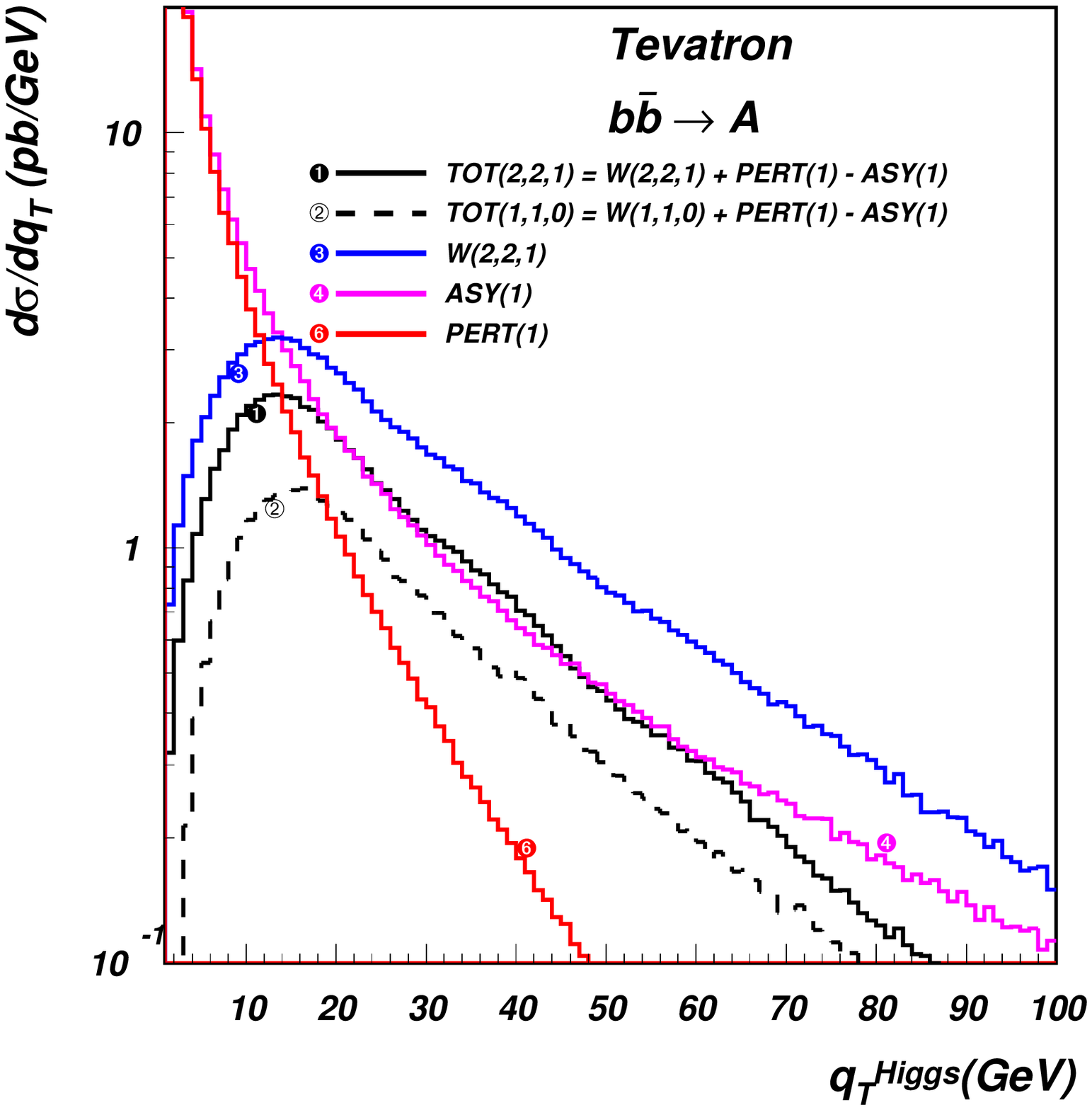}
\caption{$q_T$ distributions contributed by various pieces of the
resummation calculations. ASY(1) and PERT(1) almost 
coincide as $q_T \rightarrow 0$, and 
ASY(1) curve goes increasingly above PERT(1) curve as $q_T$ increases to 
yield a negative term Y=PERT(1)-ASY(1).}
\label{fig:two}
}
As shown in Fig.~\ref{fig:two}, 
ASY(1) and PERT(1) almost coincide as $q_T \rightarrow 0$, and then
ASY(1) curve goes increasingly above PERT(1) curve as $q_T$ increases,  yielding
a negative term Y=PERT(1)-ASY(1). 
By adding such a negative Y-term to W(2,2,1)
in TOT(2,2,1), we improve agreement of the resummation and PYTHIA predictions.
This behavior confirms our expectations from the above mentioned 
inspection of the distributions shown in Fig.\ref{fig:one}.
It is also important to know what causes the difference from the
vector boson production in light-quark annihilation channels 
({\it e.g.}, $u\bar u \rightarrow Z$), in which the Y-term is positive.
Two possible causes are the differences between the 
hard-part matrix elements included in PERT(1)
for  $b \bar b \rightarrow A$ and  
$u \bar u \rightarrow Z$ processes, and the
differences in the shapes of the parton 
distributions for $b$ and $u$ quarks at the $x$ values typical for
production of a 100 GeV Higgs boson at the Tevatron.
This will be illustrated in the next figure. 
Another comment about Fig.~\ref{fig:two} is that 
the difference in TOT(2,2,1) and TOT(1,1,0)
comes from the difference in W(2,2,1) and W(1,1,0).
 The coefficients ${\cal C}^{(1)}$ included 
in W(2,2,1) mainly change the overall
normalization to include  
(part of) the next-to-leading
order contributions that are not accounted for,
 even after
including higher-order (i.e., $O(\alpha_s^2$))
contributions ${\cal A}^{(2)}$ and ${\cal B}^{(2)}$ 
in the Sudakov form factor.
%
%
\FIGURE[htb]{
\noindent
\includegraphics[width=0.6\textwidth]{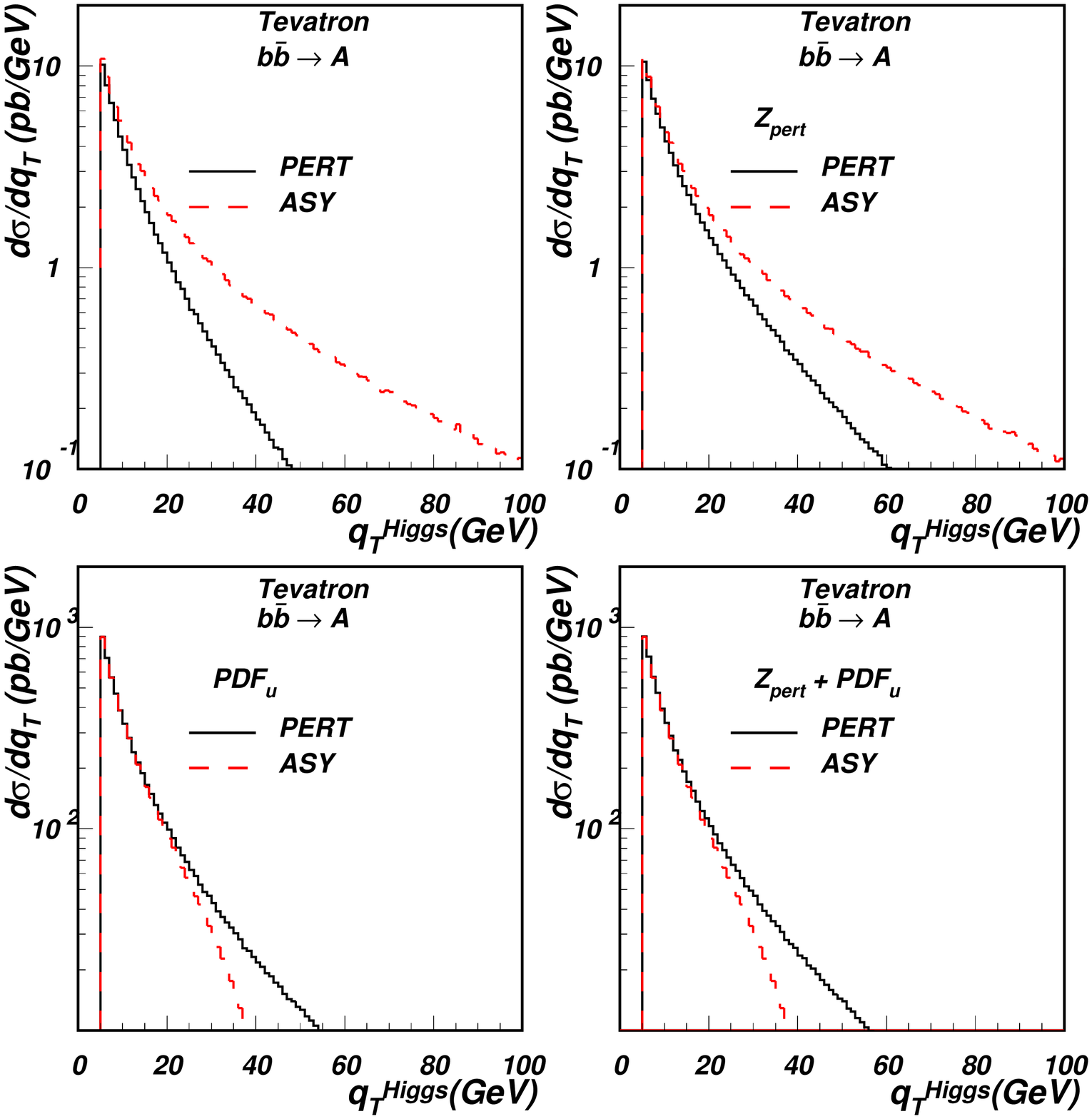}
\caption{
Sensitivity of 
$q_T$ distributions to the 
choice of hard-part matrix elements in PERT(1) and parton
distribution functions.
The upper-left plot is the reproduction of Fig.~2.
In the upper-right plot, the 
hard-part matrix element of $b {\bar b} \to A$ is replaced by 
that for $u {\bar u} \to Z$, but the PDF is not changed.
In the lower-left plot, the bottom-quark PDF is replaced  
by the up-quark PDF, but the hard-part matrix element of $b {\bar b} \to A$
is not changed. 
In the lower-right plot, both the $Z$ hard-part matrix element and $u$-quark 
PDF are used in the calculation. 
The plots show that the Y-term in $b {\bar b} \to A$ is negative mainly due to
the strong $x$ dependence of the bottom quark PDF.
}
\label{fig:three}
}

To examine the effects of the hard-part matrix elements
and the shape of
parton distribution functions on the shape of $q_T$ distribution, 
Fig.~\ref{fig:three} 
compares the results of four calculations. 
The upper-left plot shows again the result of our $b \bar b \rightarrow A$
calculation presented in Fig.~\ref{fig:two}.
The upper-right plot is obtained from the upper-left plot
by replacing the hard-part matrix element PERT(1)
for producing a CP-odd Higgs boson $A$ by  that for producing 
a vector boson $Z$. 
(Note that the quark masses are neglected in both cases.)
The overall normalization factor contributed by the $b\bar
bA$ vertex is not changed. Consequently, the difference between 
the two upper plots is 
entirely due to the kinematical dependence in the hard-part 
matrix elements in the two processes.

The lower-left plot is obtained from the upper-left plot for 
$b\bar b\rightarrow A$ by replacing the $b$  and $\bar b$ PDF's 
by $u$ and $\bar u$ PDF's.
This is to examine the effect from the different shapes of the $b$-
and $u$-quark PDF's.
Finally, the lower-right plot is obtained from the upper-left plot 
by replacing both the hard-part matrix element for 
$b\bar b \rightarrow A$
and $b$-quark PDF by the hard-part matrix element for $u\bar u
\rightarrow Z$ and $u$-quark PDF.

The four plots  clearly show that the difference 
due to the replacement of the hard-part matrix element is
small, while the difference due to the replacement of the PDF is
large. If we replace the bottom quark PDF by the up quark PDF, then the
Y-term becomes positive, in agreement with what has been observed in
the previous  calculations for vector boson production. 
From this comparison, we conclude that the shape of bottom quark 
PDF causes the Y-term to be negative.
As all sea-parton PDF's, the $b$-quark PDF is a
rapidly decreasing function of $x$ in the whole range of $x$, while
the $u$-quark PDF, which includes a substantial valence component,
varies slower with $x$.
As discussed in Section 2, the energy-momentum conservation in PERT(1)
prevents too small $x\approx Q/\sqrt{S}$ 
from contributing at nonzero $q_T$.  Meanwhile, such $x$ are included in
ASY(1), evaluated using small-$q_T$ phase space. Consequently
ASY(1) may be enhanced compared to PERT(1) by large PDF's from smaller
$x$, especially in the case of the sea-quark PDF's.
As a result of the PDF-induced enhancement in $b\bar b$ scattering, 
inclusion of the Y-term brings the resummation calculation closer to
the PYTHIA prediction. 

As discussed in Section 2, the ``kinematical correction'' (KC), which 
compensates  for small, but nonzero energy of soft gluon emissions, 
can play an important role in explaining the difference between
the resummation and PYTHIA predictions. 
Thus, we calculated ${\rm TOT}_{\rm KC}$(2,2,1) by adding ${\rm W}_{\rm KC}$(2,2,1) and
PERT(1) and subtracting ${\rm ASY}_{\rm KC}$(1) to compare it 
with TOT(2,2,1) in Fig.~\ref{fig:four}. 
It is indeed the case that the kinematical correction
reduces the rate in the large-$q_T$ region, which is consistent with
our intuition that $q_T$ will be reduced by
 the emission of (soft) gluons carrying away some energy. 
Fig.~\ref{fig:four} also shows that the 
location of the peak of the $q_T$ distribution is not affected by 
the kinematical correction. 
%
\FIGURE[htb]{
\noindent
\includegraphics[width=0.6\textwidth]{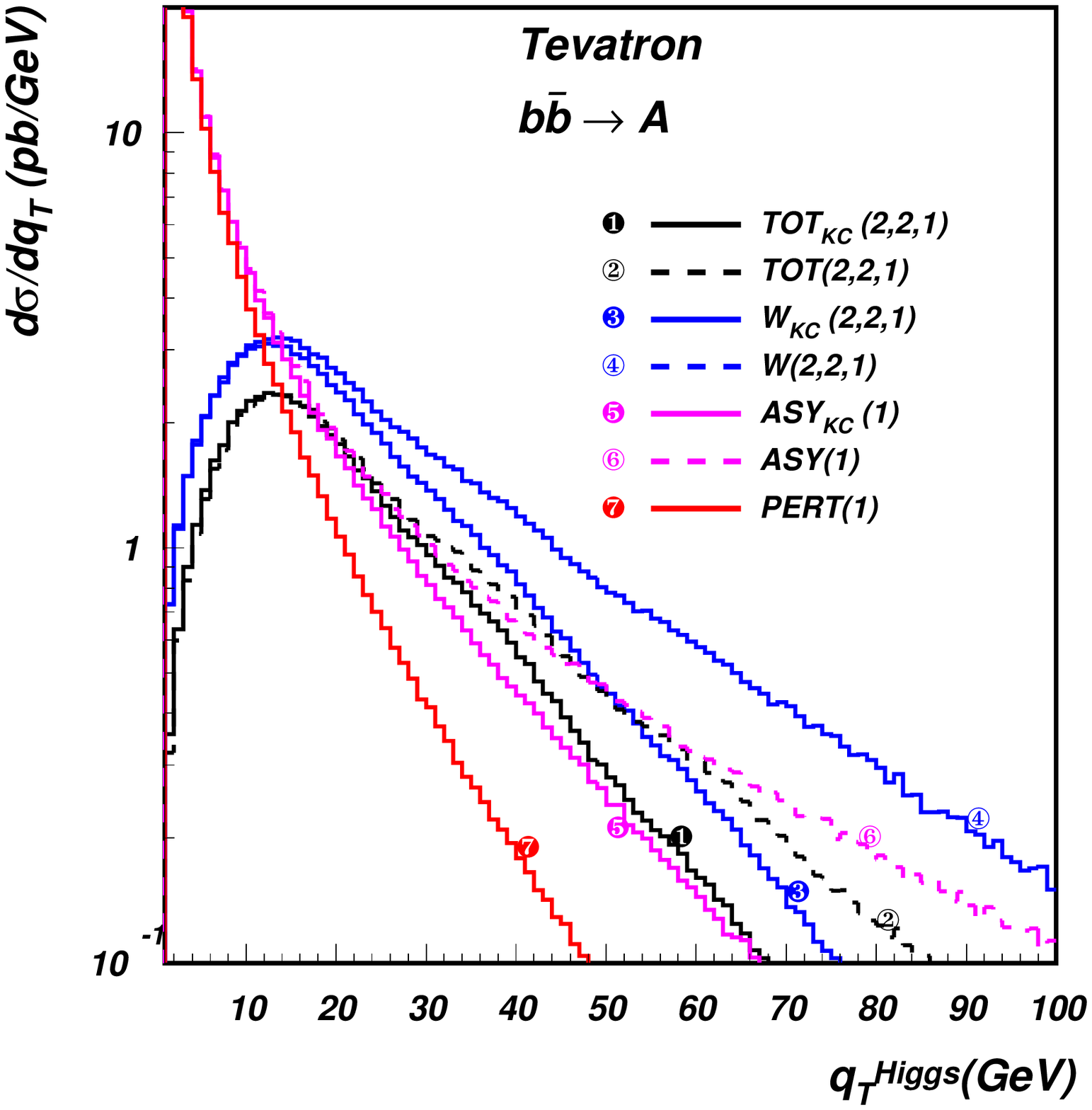}
\caption{Impact of the kinematic correction on the $q_T$
 distribution. The kinematical correction reduces the rate 
 in the large-$q_T$ region.}
\label{fig:four}
}
Hence, to explain why the peak position 
predicted by PYTHIA is lower than that in
the resummation calculations discussed above, 
we should take into account another important
piece of physics, which is included in the PYTHIA event
generator, but not in the above resummation calculations.
That is the effect of the bottom-quark mass $m_b$ on the 
unintegrated parton distribution of the bottom quark inside the resummed
W term. 
PYTHIA always generates the initial-state
bottom quarks from the gluon splittings, with the proper
kinematics for the massive $b\bar b$ pairs 
imposed at the end of the backward-radiation 
cascade chain~\cite{Sjostrand:2001yu}. 
However, the proper $m_b$ dependence was not included in the massless
resummation
calculation discussed thus far. 
The $m_b$ dependence can be introduced in the CSS resummation
formalism using the method developed 
in Ref.~\cite{Nadolsky:2002jr} and discussed in detail in Section~2.
The effect of the heavy-quark mass correction can be clearly seen from 
Fig.~\ref{fig:five}.
While the kinematical correction reduces the rate at
large $q_T$, the heavy mass $m_b$ shifts the peak of the distribution
to lower $q_T$, so that the resummation cross section 
agrees better with the PYTHIA cross section.

%
\FIGURE[htb]{
\noindent
\includegraphics[width=0.65\textwidth]{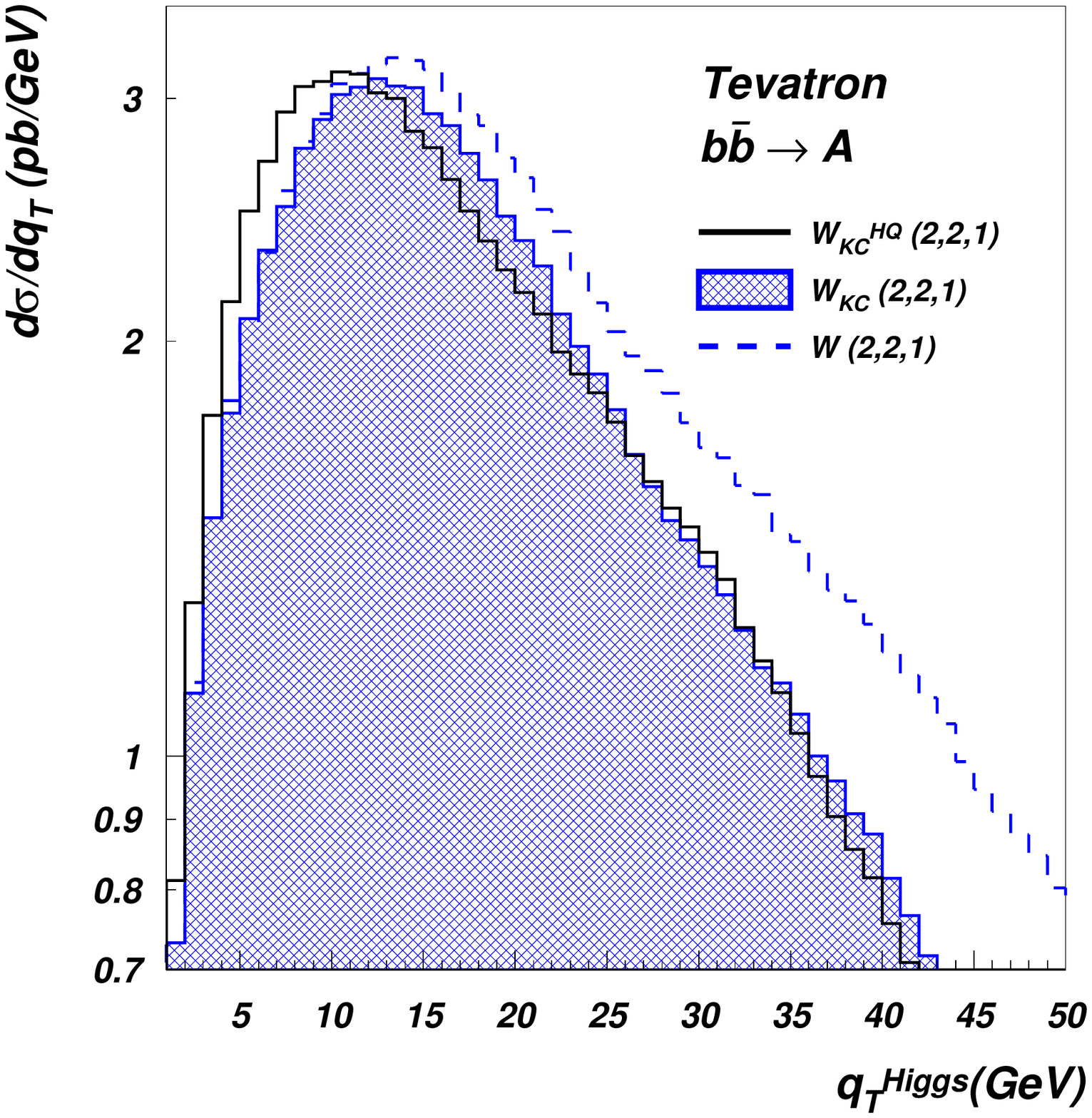}
\caption{ The combined effect of the heavy-quark mass (HQ) and
  kinematical (KC) corrections. The heavy-quark mass correction 
shifts the peak position of the 
$q_T$ distribution to a lower $q_T$, while the kinematical correction 
reduces the rate in the large-$q_T$ region. }
\label{fig:five}
}

In conclusion, our prediction on the $q_T$ distribution of Higgs boson
produced via bottom quark fusion in hadron collisions
is given by TOT(1), which is obtained 
by adding ${\rm W}_{\rm KC}^{\rm HQ}$(2,2,1) and 
PERT(1) and subtracting ${\rm ASY}_{\rm KC}$(1).
The numerical result is shown in Fig.~\ref{fig:six}~(left) 
for Tevatron ($m_A=100$~GeV) and 
in Fig.~\ref{fig:six}~(right) for LHC  ($m_A=300$~GeV), respectively,
where
TOT(1) is also compared to the PYTHIA prediction and the fixed-order
prediction PERT(1). 

As one can see, results for Tevatron and LHC are qualitatively similar.
It remains to be the case that the peak position 
in $q_T$ distribution predicted
by PYTHIA is lower than that by TOT(1), and in the 
large $q_T$ region, TOT(1) rate is larger than the PYTHIA rate. 

\FIGURE[htb]{
\noindent
\includegraphics[width=0.5\textwidth]{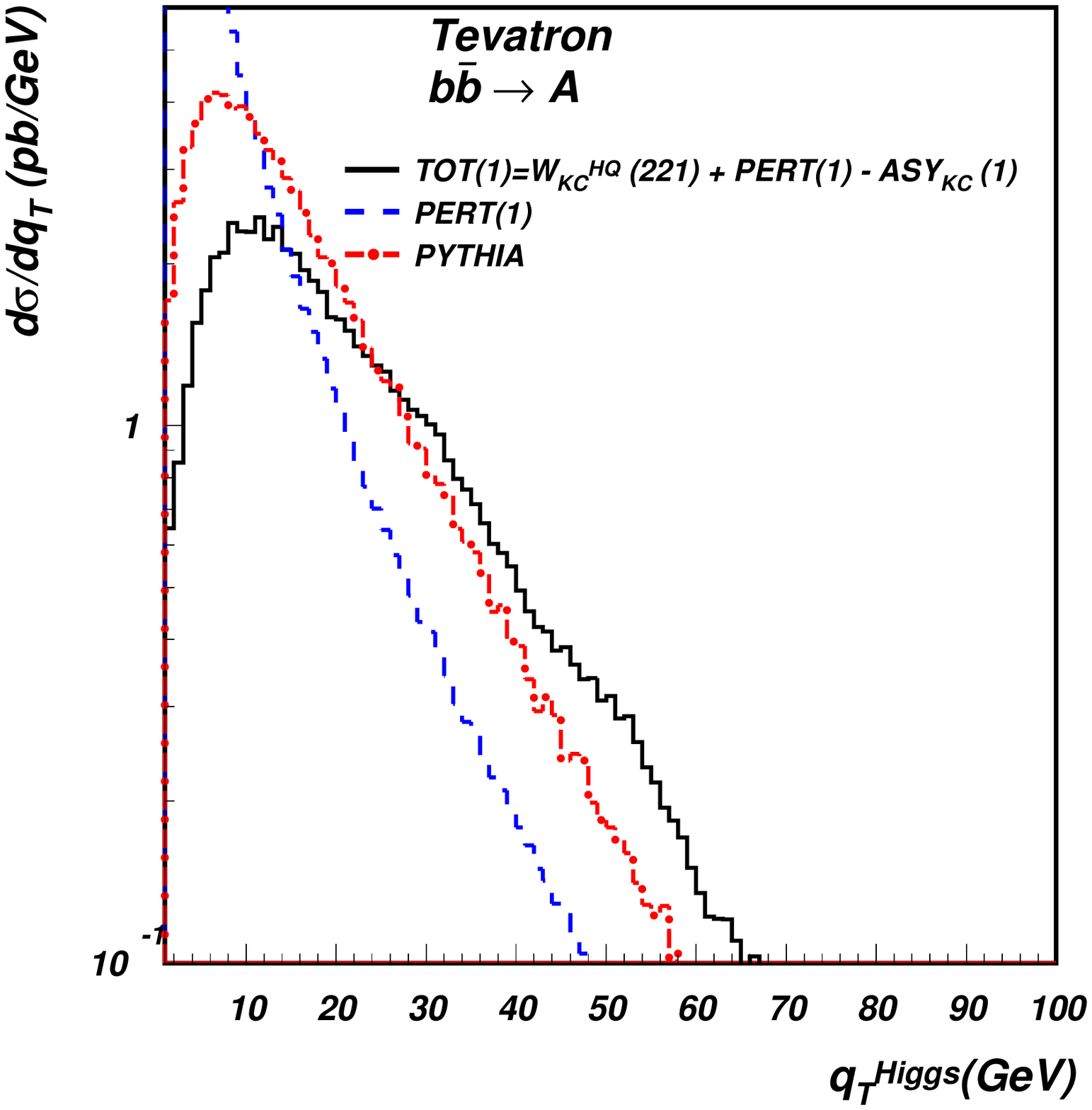}%
\includegraphics[width=0.5\textwidth]{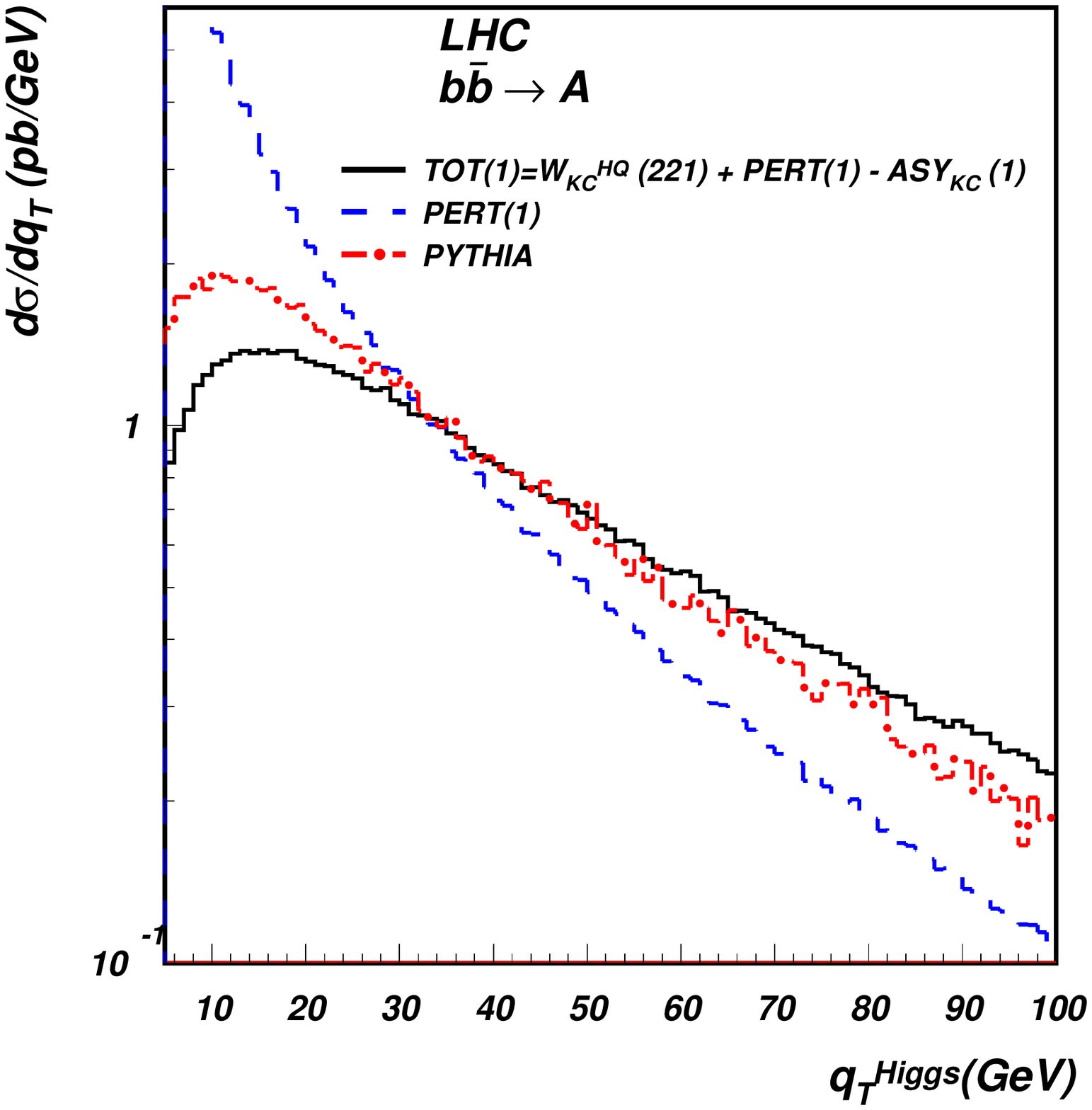}
\caption{Comparison of 
$q_T$ distributions predicted by TOT(1), PERT(1) and PYTHIA, for 
 Higgs boson produced via $b \bar b$
fusion at the Tevatron Run-2 (left) and LHC (right)
for $m_A=$100 and 300 GeV respectively.}
\label{fig:six}
}

Next, we examine how our conclusions are modified by
the ${\cal O}(\alpha_s^2)$  corrections to the Y-term. 
We calculate ${\rm ASY}_{\rm KC}(2)$ using its exact
$O(\alpha_s^2)$ expression and estimate the PERT(2) piece 
in the large-$q_T$ region 
by multiplying PERT(1) by the ${\cal O}(\alpha_s^2)$ 
``K-factor'' ($\sim 1.75$)
for the Tevatron case,
extracted from Fig.~11 in the recent calculation~\cite{Campbell:2002zm} of 
the $O(\alpha_s^2)$ rate for Higgs boson production at large $q_T$.
The resulting TOT(2), evaluated similarly to TOT(1), but using
the exact ${\rm ASY}_{\rm KC}$(2) and the estimated PERT(2), 
is only slightly smaller than 
TOT(1), not more than a few percent, for $q_T$ larger than 30 GeV
for the Tevatron case.

\FIGURE[htb]{
\noindent
\includegraphics[width=0.5\textwidth]{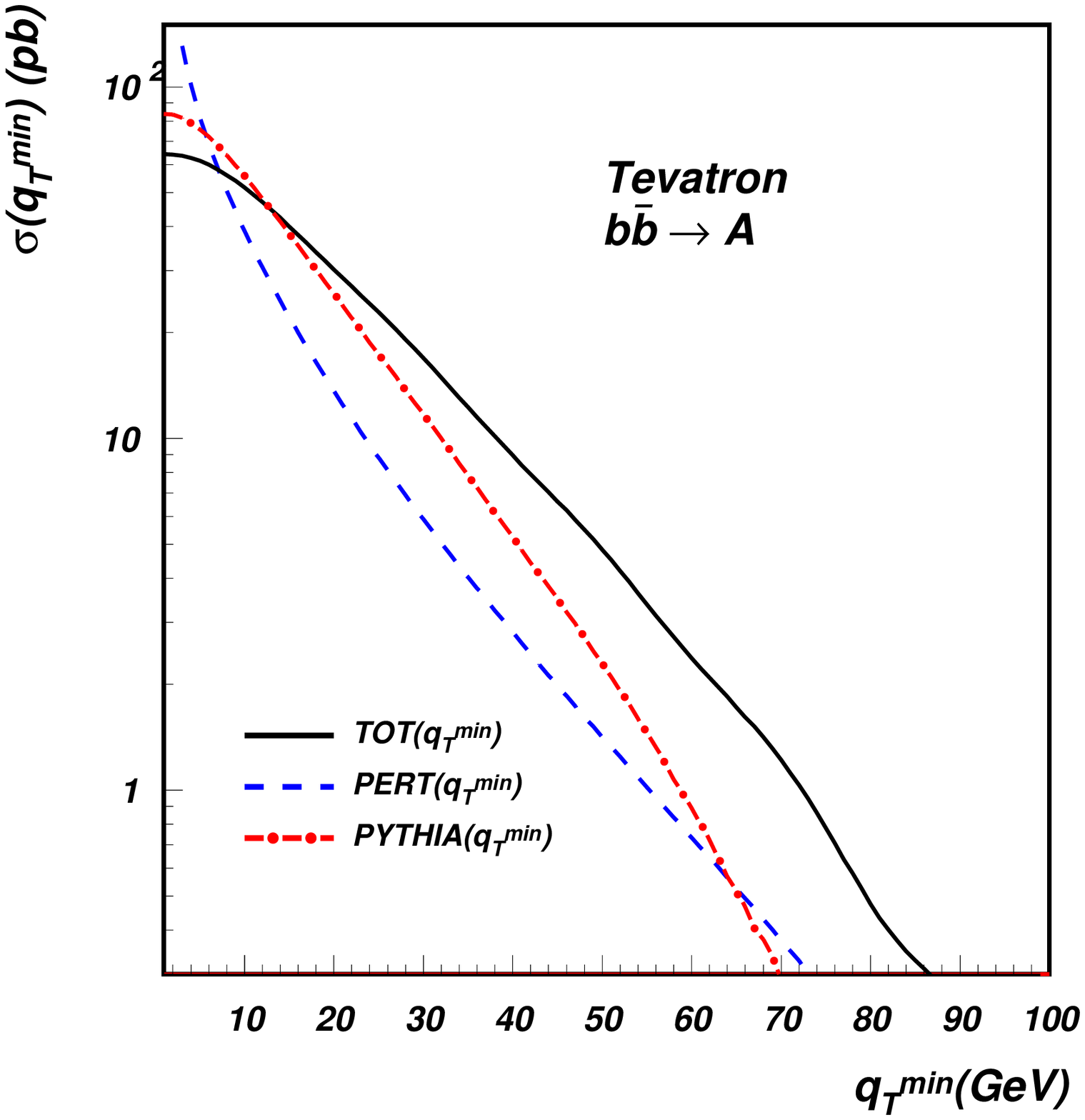}%
\includegraphics[width=0.5\textwidth]{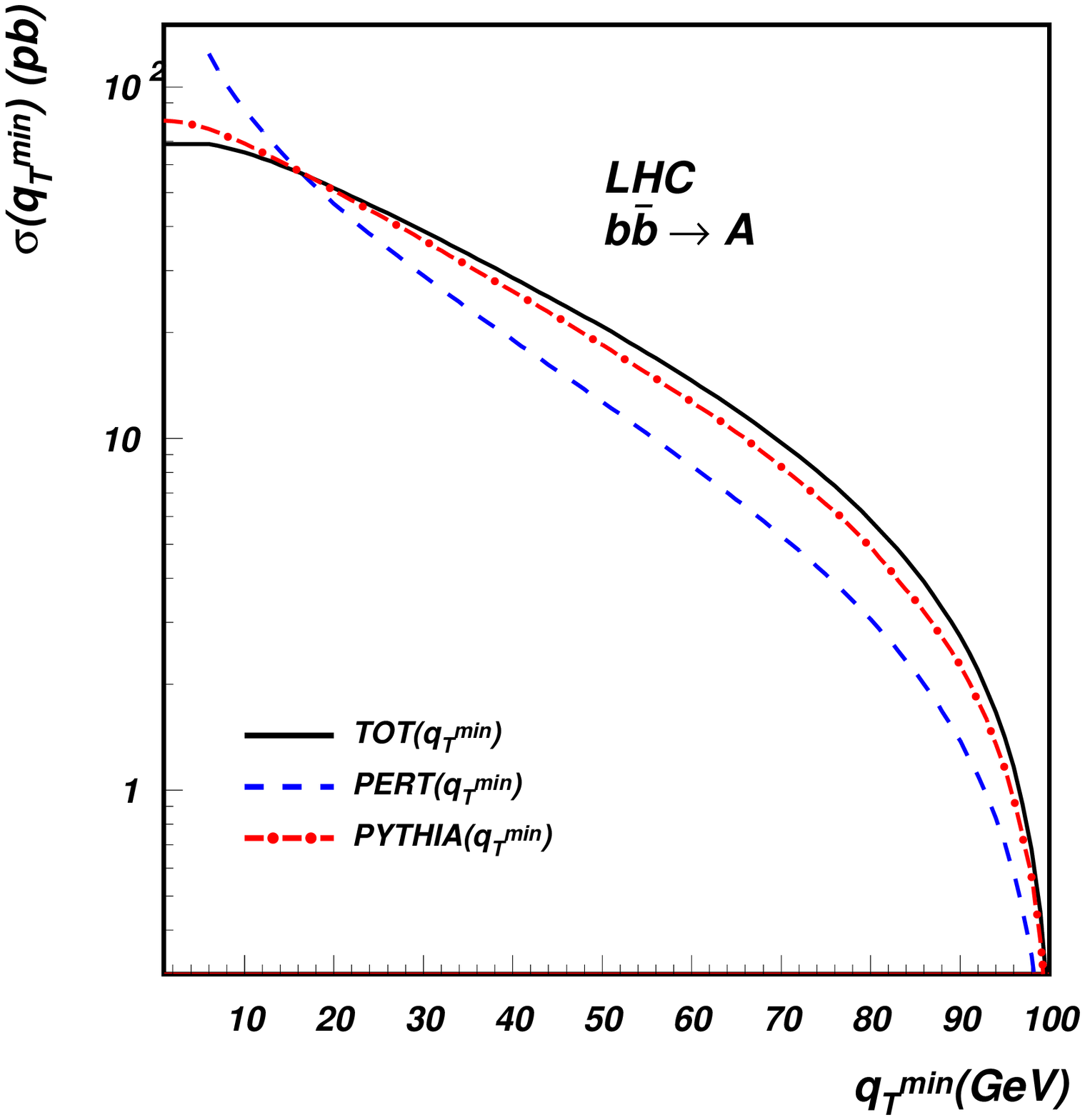}
\caption{ Comparison of the integrated rates, deduced form Fig.~6,
 as a function of the 
minimal $q_T$ value taken in the integration
at the   Tevatron Run-2 (left) and LHC (right)
for $m_A=$100 and 300 GeV, respectively.}
\label{fig:seven}
}
Finally, Fig.~\ref{fig:seven} shows the integrated cross section as
a function of the minimal $q_T$ in the calculation 
for the Tevatron (left) and LHC (right).
This is another way to illustrate the differences in the shape of 
$q_T$ distribution obtained in the resummation, fixed-order, and PYTHIA 
calculations.

\section{Discussion and Conclusion}

We studied the effect of initial-state multiple soft-gluon radiation
on the transverse momentum ($q_{T}$) distributions of Higgs boson
produced via bottom-quark fusion at hadron colliders. Due to the
shape of the bottom-quark parton distribution function that rapidly
decreases with $x$, the Y-term in the $b\bar{b}\rightarrow A$
process is negative, and the kinematical correction largely reduces
the rate in the high-$q_{T}$ region. After the $b$-quark mass is
consistently taken into account in the CSS-HQ resummation formalism
\cite{Nadolsky:2002jr}, the position of the peak in the $q_{T}$ distribution
shifts to a lower value, while the rate in the high-$q_{T}$ region
remains unchanged. The combination of the kinematic correction
 and heavy-quark
mass correction makes the resummation predictions resemble more the
PYTHIA predictions, as shown in Fig.~\ref{fig:six}.

As noted in Sec.~\ref{sec:InclXS}, we
assumed the non-perturbative functions in the CSS-HQ resummation formalism
for the heavy bottom quark are the
same as those for the light quarks in our numerical calculations.
However, additional non-perturbative dynamics, such as that associated
with {}``intrinsic'' heavy quarks \cite{Brodsky:1980pb}, might be
present in the heavy-flavor channels.
In the future, when experimental data (likely, from 
the associated production of $Z$ boson with bottom (anti)-quark
\cite{Campbell:2003dd,Mutaf:2004wc,Maltoni:2005wd}, 
and from 
the t-channel single-top production 
\cite{Dawson:1986tc,Willenbrock:1986cr,Yuan:1989tc,Carlson:1993dt,%
Heinson:1996zm,Stelzer:1997ns,Belyaev:1998dn,Stelzer:1998ni,Tait:2000sh,Cao:2005pq})
becomes available to measure the bottom quark parton distribution
function together with
those non-perturbative functions,
 we will be able to improve our theoretical
prediction on the $q_{T}$ distribution of any hard-scattering process
that is initiated by bottom quark interaction in the initial state.
Nevertheless, we expect
the qualitative comparison between TOT(1) and PYTHIA predictions shown
in Fig.~\ref{fig:six} to stay valid, as these non-perturbative variations
at impact parameters $b\gtrsim b_{0}/m_{b}\approx0.25\mbox{ GeV}^{-1}$
would result in mild modulations in $q_{T}$ space at $q_{T}\lesssim30$
GeV. More importantly, differences in the shape of the two predictions
in the large-$q_{T}$ region have implications for the discovery potential
of the Higgs boson, as they will affect the significance of the
signal event after imposing the kinematic cut on
the transverse momentum of the Higgs boson to suppress background 
or to enable the reconstruction of the signal kinematics.

\section*{Acknowledgments}

The effect of heavy-quark mass correction to the production of 
Higgs boson via $b \bar b$ fusion was 
calculated independently in Ref.~\cite{Berge:2005rv},
which was brought to the attention of A. B. and C.-P. Y. at the write-up
stage of the project. 

This work was supported in part by the U.S. National Science Foundation
under awards PHY-0354838 and PHY-0244919, and by the U.S. Department
of Energy, High Energy Physics Division, under Contract W-31-109-ENG-38.
C.-P.Y. is grateful for the hospitality of National Center for Theoretical
Sciences in Taiwan, R.O.C., where part of this work was performed.

\bibliographystyle{JHEP}   
\bibliography{hbpt}

\end{document}